\documentstyle[preprint,aps]{revtex}

\def\beq{\begin{equation}}
\def\eeq{\begin{equation}}

\skip\footins = 14pt 
\begin{document}

\nonfrenchspacing
\flushbottom

\title{Seeking an Even-Parity Mass Term for 3-D Gauge Theory\footnotemark[1]}

\footnotetext[1] {\baselineskip=16pt This work is supported in part by funds
provided by  the U.S.~Department of Energy (D.O.E.) under contracts
\#DE-FC02-94ER40818 (RJ) and \#DE-FG02-91ER40676 (SYP). \newline MIT-CTP-2615 \hfil BU-HEP-97-9
\hfil  March 1997\break}

\author{R.~Jackiw and So-Young Pi\footnotemark[2]}

\address{Center for Theoretical Physics\\ Massachusetts Institute of
Technology\\ Cambridge, MA ~02139--4307}

\footnotetext[2] {\baselineskip=16pt On sabbatical leave from Boston University, Boston, MA~~02215}

\maketitle

\vskip1in

\begin{abstract}
\noindent%
Mass-gap calculations in three-dimensional gauge theories are discussed.  Also we present a
Chern--Simons-like mass-generating mechanism which preserves parity and is realized
non-perturbatively.

\end{abstract}

\setcounter{page}{0}
\thispagestyle{empty}


\newpage

\baselineskip=24pt plus 1pt minus 1pt

Aside from their theoretical interest, three-dimensional gauge theories [on Euclidean
or $(2+1)$ Minkowski space] merit study because they describe (1) kinematical
processes that are confined to a plane when structures (magnetic fields, cosmic strings)
perpendicular to the plane are present, and (2) static properties of $(3+1)$ dimensional
systems in equilibrium with a high-temperature heat bath.  An important issue is
whether the apparently massless gauge theory possesses a  mass gap.  The suggestion
that indeed it does, gains support from the observation that the gauge coupling constant
squared, $g^2$, carries dimension of mass, thereby providing a natural mass-scale (as in
the two-dimensional Schwinger model).  Also without a mass gap, the perturbative
expansion is infrared divergent, so if the theory is to have a perturbative definition,
infrared divergences must be screened, thereby providing evidence for magnetic screening
in the four-dimensional gauge theory at high temperature.  But in spite of the above
indications, a compelling theoretical derivation of the desired result is not yet available,
even though many approaches have been tried.  These days, a popular framework for
approximately determining a mass makes use of various gauge invariant and
self-consistent gap equations.  Here we shall examine several of these
calculations, developing some of them further, and commenting on the inherent limitation of
the entire program.  We shall also describe a novel field theoretical structure that, like the
Chern-Simons paradigm\cite{ref:1}, relies on the geometric properties of 3-space to create a mass
for the gauge field, but  unlike the Chern-Simons form, it is neither parity-violating nor
perturbatively realized.

\vskip.2in

{\bf A. Gap Equations.}~~Gap equations for determining possible mass generation can be derived
in the following manner.  To the usual massless Yang-Mills action $I_{YM} (A)= -tr \int
F^\mu F_\mu$, $F^\mu
\equiv \frac{1}{2} \epsilon^{\mu \alpha \beta} F_{\alpha \beta}$, (unless otherwise noted, we
use a Euclidean formulation and contract fields with anti-Hermitian Lie algebra matrices) one adds
{\bf and} subtracts a mass action
$\Gamma^m$ -- a typically non-local but gauge invariant functional of $A_\mu$ (which could be
localized by introducing auxiliary fields).  Perturbative calculations are rearranged by defining
contributions of the {\bf subtracted}
$\Gamma^m$ to be at one loop higher than those coming from the {\bf added}
$\Gamma^m$.  This can be formalized by introducing a loop-counting parameter $\ell$,
rescaling all fields by $\sqrt{\ell}$, and calculating in a formal $\ell$-expansion with
the modified action
\begin{equation}
\Gamma_{\rm effective}=\frac{1}{\ell} \Bigg(I_{YM} (\sqrt{\ell} A) +
\Gamma^m (\sqrt{\ell} A)\Bigg) - \Gamma^m (\sqrt{\ell} A)
\label{eq:1}
\end{equation}
$\Gamma^m$ is chosen so that in tree approximation the vector field $A_\mu$ carries an even parity
mass $m$, which is determined self consistently by requiring that the transverse
portion of the momentum-space vacuum polarization tensor, computed from
(\ref{eq:1}), vanishes at $p^2=-m^2$.  Actual calculations truncate the $\ell$-power series
at first order, so the one-loop gap equation reads
\begin{equation}
\Pi^{\rm one-loop}(p^2) |_{p^2=-m^2}=m^2
\label{eq:2}
\end{equation}
where the transverse vacuum polarization tensor $\Pi^{\rm one-loop}$ is
determined by $I_{YM} + \Gamma^m$.

\vskip.2in

{\bf B.~One-Loop Calculations.}~~It remains to choose a definite expression for $\Gamma^m$ and
several possibilities present themselves.

There is the (analytically-continued) Chern-Simons eikonal used by Alexanian and Nair 
\cite{ref:2}. This non-local functional of $A_\mu$, which is a three-dimensional generalization of
the two-dimensional Polyakov--Wiegman determinant, can be localized with the help of an auxiliary
field, but this involves passing to a (fictitious) fourth dimension and integrating on the
three-dimensional boundary of the four-dimensional space.  In Feynman gauge for the $SU(N)$ gauge
theory the vacuum polarization tensor is transverse, and the invariant function is\cite{ref:2}
\begin{equation}
\Pi^{\rm one-loop}_{A \cdot N} (p^2) = {Ng^2m \over 8\pi} \left[ 
\left(
-\frac{5}{4} \frac{p^2}{m^2} + 4 \right) \frac{2m}{p} \tan^{-1} \frac{p}{2m} -
\frac{p^2}{m^2} \left( 1+ \frac{m^2}{p^2}\right)^2 \frac{m}{p} \tan^{-1} \frac{p}{m} +
\frac{m^2}{p^2} \right]
\label{eq:3}
\end{equation}

An alternative form for $\Gamma^m$ is suggested by the two-dimensional structures encountered in
the Schwinger model and in Polyakov's induced gravity action: curvature $\times$ inverse
invariant Laplacian $\times$ curvature.  Thus we have chosen\cite{ref:3}\footnote{We take this
opportunity to acknowledge mis-spelling ``threshold'' throughout Ref.~\cite{ref:3}}
\begin{equation}
\Gamma^m_{J \cdot P}(A) = m^2tr \int F^\mu \frac{1}{D^2} F_\mu
\label{eq:4}
\end{equation}
$$
D^2=D^\mu D_\mu
$$
where $D_\mu$ is the gauge covariant derivative.  The resulting transverse vacuum
polarization, again in Feynman gauge, is 
\begin{eqnarray}
\Pi^{\rm one-loop}_{J \cdot P}(p^2) &= & {Ng^2m \over 8\pi} \left[ 
\left(
\frac{1}{16} \frac{p^6}{m^6} + \frac{1}{4} \frac{p^4}{m^4} - \frac{3p^2}{4m^2} +
\frac{47}{8} - \frac{1}{2} \frac{m^2}{p^2} 
\right) \frac{2m}{p} \tan^{-1} \frac{p}{2m}  \right. \nonumber \\ 
&& \quad - \left(
\frac{1}{4} \frac{p^6}{m^6} + \frac{1}{2} \frac{p^4}{m^4} - \frac{1}{2} \frac{p^2}{m^2} +
\frac{1}{4}
\right) \left(1 + \frac{m^2}{p^2} \right)^2 \frac{m}{p} \tan^{-1} \frac{p}{m}  \nonumber \\
 && \quad -\frac{p^2}{4m^2} - 2 + \frac{49}{12} \frac{m^2}{p^2} + \frac{1}{4} \frac{m^4}{p^4} + 
\left. \left(
\frac{1}{16} \frac{p^4}{m^4} + \frac{1}{4} \frac{p^2}{m^2} - \frac{5}{8}
\right) \pi  \frac{p}{m}
\right]
\label{eq:5}
\end{eqnarray}

It is instructive to examine the analyticity properties of (\ref{eq:4}) and (\ref{eq:5}), which are
presented at Euclidean momenta, but need to be evaluated in the gap equation at the
Minkowski value $p^2 = -m^2$, reached by analytic continuation of the inverse
tangent according to $\frac{1}{x} \tan^{-1} x = \frac{1}{2 \sqrt{-x^2}} \ln
\frac{1+\sqrt{-x^2}}{1-\sqrt{-x^2}}$.

Evidently the singularity at $p^2 = -4m^2$ (in $\frac{2m}{p} \tan^{-1} \frac{p}{2m}$) is a
threshold arising from the exchange (emission) of two gauge propagators, each with mass
$m$.  There is also a singularity at $p^2 = -m^2$ (in $\frac{m}{p} \tan^{-1}
\frac{p}{m})$, which is understood as follows.  Even though the boson propagators
are massive, the non-local vertices in $\Gamma_{A \cdot N}^m$ and $\Gamma_{J \cdot P}^m$ contain
$1/({\rm momentum})^2$ factors, and these provide massless lines.  Therefore
the singularity at $p^2 = -m^2$ may be viewed as a threshold for the emission of a massive
boson propagator and a massless line from the vertex.   Finally in $\Pi^{\rm one-loop}_{J \cdot P}
(p^2)$ there are  singularities at $p^2=0$ (in $\frac{p}{m}$), which again are interpreted as
thresholds for massless vertex lines.  [Individual graphs contributing to $\Pi^{\rm one-loop}_{A
\cdot N} (p^2)$ similarly contain singularities at $p^2=0$, but these cancel in the sum.]  Only the
last two thresholds  have an effect on the gap equation; the one at $p^2 = -4m^2$ is irrelevant to
the point of interest $p^2 = -m^2$.

The interpretation of thresholds in terms of massless exchange can be substantiated for 
$\Gamma^m_{J \cdot P}$ in the following fashion.\cite{ref:4}  Expression~(\ref{eq:4}) can be
localized with the help of a multiplet of auxiliary vector fields $\phi_\mu$ in the adjoint
representation, governed by the action
\begin{equation}
\Gamma^m_{J \cdot P}(A,\phi) = -tr \int \left[ (D_\mu \phi_\nu) (D^\mu \phi^\nu) + 
2im  F^\mu \phi_\mu \right]
\label{eq:6}
\end{equation}
When $\phi_\mu$ is functionally integrated one arrives at $\Gamma^m_{J \cdot P}(A)$, (provided the
$\phi_\mu$ measure includes
$\det^{\frac{1}{2}}D^2$).  Note the presence of a Chern-Simons-like term; the factor
$i=\sqrt{-1}$ in Euclidean space corresponds to a real expression in Minkowski space. 
While the mixing term gives rise to a mass for some of the fields, the (covariantly)
longitudinal portion of $\phi_\mu$, which is present in the kinetic part of (\ref{eq:6}), does not
participate in the mass-generating interaction, owing to the Bianchi identity satisfied by
$F^\mu$.  Consequently that component propagates as a massless field and
presumably is responsible for the massless thresholds in $\Pi^{\rm one-loop}_{J \cdot P}$. 
[Since dimensional extension is needed to localize $\Gamma_{A \cdot N}^m$, the reason for the
massless threshold in $\Pi^{\rm one-loop}_{A \cdot N}$ is not so evident, but recall that massless
ghosts certainly contribute to Eq.~(\ref{eq:3}).]  

Below we shall suggest a modification of Eq.~(\ref{eq:6}), for which the (covariantly)
longitudinal components of $\phi_\mu$ decouple, and satisfactory non-perturbative mass
generation is achieved.

A third choice for $\Gamma^m$ has been analyzed by Buchm\"uller and Philipsen\cite{ref:5}, who
make use of the non-linear $\sigma$-model, with Goldstone fields $\Phi = m U$ where 
$U$ is a unitary matrix (group element), $U^\dagger U = I$.\footnote{In a related context the
mass-generating features of $\sigma$-models were used earlier by J. Cornwall.~\cite{ref:6}}
\begin{equation}
\Gamma^m_{B \cdot P} = tr \int  (\partial_\mu \Phi + A_\mu \Phi)^\dagger (\partial_\mu \Phi +
A_\mu \Phi) 
\label{eq:7}
\end{equation}
Calculation is performed in the $R_\xi$ gauges (where the vacuum polarization is {\bf not}
transverse) and in the Feynman, $\xi = 1$, gauge one finds for the transverse component
\begin{equation}
\Pi^{\rm one-loop}_{B \cdot P}(p^2) = {Ng^2m \over 8\pi} \left[ 
\left(
-\frac{27}{16} \frac{p^2}{m^2} + \frac{9}{4} \right) \frac{2m}{p} \tan^{-1} \frac{p}{2m} -
\frac{3}{4} \right]
\label{eq:8}
\end{equation}
This expression has only the threshold at $p^2 = -4m^2$; the other dangerous
thresholds at $p^2 = -m^2,0$ are absent, because in $R_\xi$ gauges with $\xi>0$,  the
Goldstone and ghost fields are both massive.

The mass as determined by the gap equations is, in the three
cases\footnotemark[3]

\footnotetext[3]{\baselineskip=16pt  Owing to a typesetting error, ``$\ln 3$'' is missing from
the corresponding formula for $m_{J \cdot P}$ in Ref.~[3]}
\begin{equation}
m_{A \cdot N}  = {Ng^2 \over 8\pi} 
\left[ 
\frac{21}{4} \ln 3 -1 \right]
\label{eq:9}
\end{equation}
\begin{equation}
m_{J \cdot P}  = {Ng^2 \over 8\pi} 
\left\{
\left[ 
\frac{117}{16} \ln 3 - \frac{67}{12} \right] \pm i\pi \frac{13}{16} \right\}
\label{eq:10}
\end{equation}
\begin{equation}
m_{B \cdot P} = {Ng^2 \over 8\pi} 
\left[ 
\frac{63}{16} \ln 3 - \frac{3}{4} \right]
\label{eq:11}
\end{equation}

Because of the spread in values, no definite conclusion can be drawn.  Indeed the
$m_{J\cdot P}$ expression would indicate that a mass is {\bf not} generated, since a complex
gap is found. This happens because, although the factor $(p^2 +m^2)^2$ extinguishes the thresholds
at $p^2=-m^2$ in  $\Pi^{\rm one-loop}_{A \cdot N}$ and $\Pi^{\rm one-loop}_{J\cdot P}$, the 
$p^2=0$ threshold in the latter survives, rendering that amplitude complex for $p^2<0$.

It is
remarkable that $m_{B\cdot P}$ is precisely $\frac{3}{4}m_{A \cdot N}$, but we have no
understanding of this numerical coincidence -- it does not hold off shell.  [One may
calculate $\Pi^{\rm one-loop}_{B \cdot P}$ with $\xi=0$,\cite{ref:7} where both Goldstone and
ghost fields are massless, and thresholds arising from these massless lines are present, but this
off-shell formula also bears no relation to $\Pi_{A \cdot N}^{\rm one-loop}$, though of course on
shell it reproduces (\ref{eq:11}).]  

While $m_{A \cdot N}$ and $m_{B\cdot P}$ give reasonable answers, lack of
agreement between the two, as well as the complex value for
$m_{J\cdot P}$, expose the unreliability of one-loop calculations.  But consideration
of higher loops poses further problems.  With the non-local actions $\Gamma^m_{A \cdot N}$
and $\Gamma^m_{J\cdot P}$ one is overwhelmed by the proliferation of graphs, and it
is unclear whether already at the two-loop level unfavorable thresholds in $\Pi_{A \cdot N}$
(which are absent or extinguished in the one-loop calculation) render the mass
complex.  Also there is the possibility of infinities.  These are absent at one-loop due to
the special features of three-(more generally, odd-) dimensional integration, but
presumably infinities arise at higher loops, and need to be renormalized, while retaining a
meaningful finite value for the gap.  Moreover, we shall argue below that, independent
of these technical difficulties, no reliable estimate for the gap can be found in a finite-order
loop calculation.

{\bf C.~Higher Loops}~~Calculations based on the non-linear $\sigma$-model (\ref{eq:7}) can be
reorganized and simplified.  Consider the functional integral for the partition function, with the
Goldstone field $\Phi=mU$, parametrized {\it e.g.\/} as $m \exp \phi$, with $\phi$ in the
adjoint representation of the Lie algebra.
\begin{equation}
Z=\int {\cal D} A_\mu {\cal D}\phi \triangle \exp - \frac{1}{\ell} \left(I_{YM} + I_\sigma -
\ell I_\sigma + I_g\right)
\label{eq:12}
\end{equation}
Here $I_\sigma$ is the $\sigma$-model action (\ref{eq:7})	 and fields are rescaled with the
loop-counting parameter.  $I_g$ is some gauge fixing and $\triangle$ is the
Faddeev-Popov compensator -- we do not specify these explicitly.  We propose
integrating the Goldstone field:  choose the gauge fixing to depend only on $A_\mu$
and change variables according to
\begin{equation}
A_\mu \rightarrow UA_\mu U^{-1} - \partial_\mu U U^{-1} \equiv A^U_\mu
\label{eq:13}
\end{equation}
This change of variables is like a gauge transformation, which leaves $I_{YM}$ and
$\triangle$ unchanged, since they are gauge invariant, while  $I_\sigma$ becomes $\Gamma^m=-m^2 tr
\int A_\mu A^{\mu}$.
Finally, from the definition of the Faddeev-Popov compensator, we know that $\int {\cal D} \phi
\triangle e^{-\frac{1}{\ell} I_g(A^U)}=1$.
[This is true provided terms proportional to $\delta(0)$ are ignored, as is justified in
dimensional regularization; see Ref.~8.]  Thus we are left with
\begin{equation}
Z=\int {\cal D} A_\mu \exp - \left(\frac{1}{\ell} I_{YM} - m^2 tr  \int A^2 +
\ell m^2 tr \int A^2 \right)
\label{eq:16}
\end{equation}
{\it i.e.\/}~{\bf massive} Yang-Mills theory and a subtraction term that contributes only to
one loop.

Observe that our final expression (\ref{eq:16}) can also be viewed as arising from
$\Gamma^m_{B\cdot P}$ in the unitary gauge.  However, we prefer the point of view that it results
from integrating the gauge degrees of freedom in an arbitrary gauge, rather than fixing the
gauge to be the unitary one -- a point of view which has been expressed previously.\cite{ref:9}

The representation (\ref{eq:16}) enjoys the advantage that the added and subtracted terms
have a clear and simple meaning:  they provide a mass term and only a mass term, without further
interactions, such as those in $\Gamma^m_{A \cdot N}$, $\Gamma^m_{J \cdot P}$ and $\Gamma^m_{B
\cdot P}$.
Indeed the {\bf exact} gap equation may now be simply stated:
\begin{equation}
\Pi(p^2)|_{p^2=-m^2} = m^2
\label{eq:17}
\end{equation}
where $\Pi(p^2)$ is the coefficient function in the polarization tensor (necessarily
transverse) calculated in massive Yang-Mills theory.  Consequently it is clear that all
thresholds are at or beyond $p^2=-4m^2$.  Such a gap equation is what one would
expect in a theory without gauge symmetry, where a mass term, quadratic in the relevant fields, is
added and subtracted.  Here it emerges in gauge theory, and consistency with gauge invariance
has been maintained; yet there is a noteworthy difference.  $\Gamma^m_{A \cdot N}$ and
$\Gamma^m_{J\cdot P}$ share with $\Gamma^m_{B\cdot P}$ the feature that they can be presented as
local, gauge invariant expressions, involving an auxiliary field; however, when this auxiliary
field is integrated, $\Gamma^m_{A \cdot N}$ and $\Gamma^m_{J\cdot P}$ leave non-local, but still
gauge invariant formulas, while $\Gamma^m_{B\cdot P}$ leads to $-m^2 tr \int A^\mu A_\mu$, which is
local, but not gauge invariant.

We have performed the one loop calculation within the formalism suggested here,
{\it i.e.\/} within massive Yang-Mills theory.  We find, as expected, a transverse polarization,
with
\begin{equation}
\Pi^{\rm one-loop} (p^2) = \frac{Ng^2m}{8\pi} \left\{\left( \frac{1}{16} \frac{p^6}{m^6} -
\frac{1}{2} \frac{p^4}{m^4} - \frac{5}{2} \frac{p^2}{m^2} + 2 \right) \frac{2m}{p}
\tan^{-1}  \frac{p}{2m} + \frac{1}{4} \frac{p^4}{m^4} - \frac{p^2}{m^2} -2 \right\}
\label{eq:18}
\end{equation}
and at $p^2=-m^2$ we regain (\ref{eq:11}).  This is to be expected since the above ``unitary"
calculation can be reached from the $R_\xi$ gauges with $\xi \rightarrow \infty$, but
at $p^2=-m^2$ there is no $\xi$ dependence.  One can also consider $\Pi^{\rm
one-loop}_{B\cdot P}$ for arbitrary $\xi$\cite{ref:7} and check whether (\ref{eq:18}) is regained at
$\xi\rightarrow\infty$.  We have done this; the expected agreement is verified,
provided, as has been shown in other contexts,\cite{ref:9} the $\xi\rightarrow\infty$ limit is
taken at fixed cut-off, {\it i.e.\/} before the diverging integrals are evaluated. 

With the compact and exact gap equation (\ref{eq:17}), one can appreciate the futility of
finite-loop calculations.  On dimensional grounds $\Pi(p^2)|_{p^2=-m^2}$ has the form
$m^2f\left(\frac{g^2}{m}\right)$, where $f$ is a numerical function of its argument, with
a power series corresponding to the loop expansion.  The gap equation requires setting
$f$ to unity at a specific value for $m$, which on dimensional grounds must be
proportional to $g^2$.  In other words, if we define $m=g^2\epsilon$, where $\epsilon$ is
a number, the gap equation requires $f\left(\frac{1}{\epsilon}\right)=1$, for real,
positive $\epsilon$.  On the other hand the loop expansion gives $f(x)=f_1 x + f_2 x^2 + \cdots$.
At one loop level, a solution is found as long as $f_1$ is real and positive: 
$\epsilon-f_1=0$.  But at two loops, one needs to solve $\epsilon^2-\epsilon f_1 - f_2=0$,
and existence of a solution depends on properties of $f_2$, which is in no way
``negligible" since it is a numerical quantity.  And the story continues with higher loops.

{\bf D. ~Another Mass-generating Model.}~~We construct an improved
version of $\Gamma_{J \cdot P}^m$, such that the Yang--Mills fields acquire a mass, but the
(covariantly) longitudinal vector fields decouple. ($\Gamma^m_{J \cdot P}$ is not an acceptable
dynamical action for vector mesons.)  We propose the Lagrange density (henceforth, formulas are in
Minkowski space-time)
\begin{eqnarray}
{\cal L} &=& tr (F^\mu F_\mu + G^\mu G_\mu - 2mF^\mu \phi_\mu) \nonumber \\
G_{\mu\nu} &\equiv& D_\mu \phi_\nu - D_\nu \phi_\mu, \quad G^\mu = \frac{1}{2} \epsilon^{\mu \alpha
\beta} G_{\alpha \beta} = \epsilon^{\mu \alpha \beta} D_\alpha \phi_\beta
\label{eq:19}
\end{eqnarray}
By declaring $\phi_\mu$ to carry odd parity, the model is parity conserving.  Mass generation
is established by looking to the quadratic part of ${\cal L}$.  Upon forming the linear
combinations $ \varphi^\pm_\mu  = \frac{1}{\sqrt{2}} (A_\mu \pm \phi_\mu)$,
and Abelian field strengths $f^\pm_{\mu\nu}=\epsilon_{\mu\nu \alpha} f^{\alpha \pm} =
\partial_\mu \varphi^\pm_\nu - \partial_\nu \varphi^\pm_\mu$, we can rewrite the quadratic part of
${\cal L}$, apart from a total derivative, as 
\begin{equation}
{\cal L}^{\rm quadratic} = tr (f^{\mu +} f_\mu^+ +  f^{\mu -} f_\mu^- - m f^{\mu +} 
\varphi_\mu^+ + m f^{\mu -} \varphi_\mu^-)
\label{eq:20}
\end{equation}
This describes two topologically massive gauge theories, and parity is conserved by field
interchange.\cite{ref:1}  The nonlinear Euler--Lagrange equations from (\ref{eq:19}), which are
first-order for $F$ and $G$
\begin{mathletters}
\begin{equation}
\epsilon_{\mu \alpha \beta} D^\alpha F^\beta - m G_\mu +  \epsilon_{\mu \alpha \beta}
[G^\alpha, \phi^\beta] = 0 \label{eq:21a}
\end{equation}
\begin{equation}
\epsilon_{\mu \alpha \beta} D^\alpha G^\beta - m F_\mu = 0 \label{eq:21b}
\end{equation}
may be combined into a second order equation
\begin{equation}
D^2G_\mu - D^\nu D_\mu G_\nu + m^2G_\mu - m \epsilon_{\mu \alpha \beta} [G^\alpha,
\phi^\beta] = 0 \label{eq:21c}
\end{equation}
\end{mathletters}
whose linear part again exhibits mass generation for $G$, and also for $F$ through (\ref{eq:21b}).

The full non-linear theory possesses an interesting symmetry structure.  In addition to the gauge
symmetry
\begin{equation}
\delta_1 A_\mu  =  D_\mu \theta \ \ , \qquad \delta_1 \phi_\mu  =  [\phi_\mu,\theta]
\label{eq:21}
\end{equation}
the last mixing term in ${\cal L}$ also is invariant against
\begin{equation}
\delta_2  A_\mu  =  0 \ \ , \qquad \delta_2 \phi_\mu  =  D_\mu \chi 
\label{eq:22}
\end{equation}
since $F^\mu$ satisfies the Bianchi identity.  But the second transformation does not leave the
non-linear part of $G^\mu$ invariant, because $\delta G^\mu = [F^\mu,\chi]$.  In other words, the
quadratic theory possess two independent, Abelian gauge symmetries; with interaction, one
non-Abelian symmetry survives.  This presents an intricate quantization problem.

On the basis of the operative gauge symmetry (\ref{eq:21}) in the model, one expects
that the measure in a functional integral acquires just the gauge fixing and gauge
compensating determinant relevant to (\ref{eq:21}).  That this is indeed correct emerges
after a detailed analysis, see below.

The Lagrangian (\ref{eq:19}) describes ``charged vector mesons" $\phi_\mu$ interacting
minimally with a gauge potential $A_\mu$.  A common approach is to supplement this with
additional non-minimal interactions so that (extended versions of) both gauge
transformations (\ref{eq:21}), (\ref{eq:22}) are incorporated in a larger non-Abelian
Yang-Mills gauge symmetry \cite{ref:10}.  However, it is possible to combine both
transformations (\ref{eq:21}), (\ref{eq:22}) into a non-Abelian gauge symmetry
without changing the dynamics, but the result does not follow the Yang-Mills paradigm.
To this end we introduce an additional scalar field multiplet $\rho$, which transforms under
(\ref{eq:21}) in the adjoint representation,
\begin{equation}
\delta_1 \rho = [\rho,\theta]
\label{eq:23}
\end{equation}
while the transformation (\ref{eq:22}) effects a shift
\begin{equation}
\delta_2 \rho = - \chi
\label{eq:24}
\end{equation}
Also we modify the Lagrange density (\ref{eq:19}) by adding in the kinetic term $[F^\mu, \rho]$
to
$G^\mu$, which is equivalent to working with the invariant combination $\phi_\mu + D_\mu \rho$,
and it follows that $G^\mu + [F^\mu,
\rho]$ is invariant.\footnote{This approach was developed in conversations with L. Griguolo, P.
Maraner and D. Seminara, who suggested it for a similar $U(1)$-based model that they are
investigating.  The additional $\rho$ field is analogous to the ``Poincar\'e coordinate,'' which is
used to gauge the Poincar\'e group in gravity theories.\cite{ref:11}}

Evidently we have introduced an Abelian group with as many parameters as the original gauge
group that is responsible for (\ref{eq:21}), and combined the two in a semi-direct product.  If
the non-Abelian generators are $Q_a$, and the Abelian ones are $P_a$, the Lie algebra
is 
\begin{equation}
[Q_a, Q_b] = f_{abc} Q_c \ \ , \qquad
[Q_a, P_b] = f_{abc} P_c \ \ , \qquad
[P_a, P_b] = 0 
\label{eq:24a}
\end{equation}
Associated with $Q_a$ are the gauge connection components (labeled by a) $A_\mu^a$;  and with
$P_a$, $\phi_\mu^a$.  The transformations (\ref{eq:21}) and (\ref{eq:22}) follow these definitions,
while the total curvature has $F_{\mu\nu}^a$ as its component along $Q_a$ and $G_{\mu\nu}^a$ along
$P_a$.  The Lagrange density
\begin{equation}
{\cal L}_\rho = - \frac{1}{4} F_{\mu\nu}^a F^{a\mu\nu} - \frac{1}{4} (G_{\mu\nu}^a + f_{abc}
F_{\mu\nu}^b \rho^c) (G^{a\mu\nu} + f_{abc} F^{b\mu\nu} \rho^c) + \frac{m}{2}
\epsilon^{\alpha \beta \gamma} F_{\alpha \beta}^a \phi_{\gamma}^a
\label{eq:24b}
\end{equation}
is invariant, but not of the Yang-Mills form.  At the same time, by (\ref{eq:24}) one can always
set
$\rho$ to zero, thereby regaining the dynamics (\ref{eq:19}).

The presence of the gauge symmetry allows straightforward quantization, following familiar
principles of Hamiltonian reduction\cite{ref:12}.  One arrives at a phase space functional integral
for unconstrained degrees of freedom, and integrates over the canonical momenta to regain a
configuration space functional integral.  The result in  the gauge $\rho=0$ is as anticipated
above, except that at an intermediate stage there arises the additional factor
$$
\int {\cal D} \Pi_a^i \,\, {\rm det}^{\frac{1}{2}} 
\left( f_{ace} f_{edb} (\Pi_c^i \Pi_d^i + F_c^0 F_d^0) \right) \exp -i \frac{1}{2} \left( \Pi_a^i -
\epsilon^{ij}  (G^j_a + f_{abc} \rho^b F_c^i) \right)^2 
$$
where $\Pi_a^i$ is the momentum conjugate to $\phi_i^a$.  When the determinant is written as
$\exp \frac{1}{2} tr \ln$, the argument of the trace is a
local function, so the exponent acquires a $\delta (0)$ factor, which is ignored in dimensional
regularization.  Then the $\Pi_a^i$ integral is Gaussian and irrelevant, and one is left with the
previously described, naive result.

Nevertheless, straightforward perturbation theory cannot be carried out.  This is because the
gauge-fixed Lagrangian ($\rho=0$ and appropriate gauge fixing for $A_\mu$) while non-singular in
its entirety, possess a singular term quadratic in $\phi_\mu$, so a propagator cannot be
defined, unless one expands around a non-trivial background for $A_\mu$.  This is also seen in
the above (ignorable) factor $(\Pi_c^i \Pi_d^i + F_c^0 F_d^0)$, which does not possess an
expansion around vanishing argument.  Yet another version of the same story is seen after
integrating (\ref{eq:19}) over
$\phi_\mu$:
\begin{equation}
Z = \int {\cal D}\mu (A) \,\, {\rm det}^{\frac{1}{2}}  M \exp i [I_{YM} - \frac{1}{2} m^2 \int
F_a^\mu M_{\mu\nu}^{ab} F_b^\nu]
\label{eq:27}
\end{equation}
${\cal D}\mu (A)$ is an ordinary gauge fixed and compensated Yang--Mills measure and
$M_{\mu\nu}$ is the inverse of $D^2 g_{\mu\nu} - D_{(\mu} D_{\nu)}$.  But the
inverse does not exist when the covariant derivatives are replaced by ordinary derivatives.  So the
inverse can be constructed only by retaining a background for $A_\mu$.  This, together with
the fact (\ref{eq:24}) that the $\rho$ field transforms by a $c$-number shift, suggests
spontaneous symmetry breaking.  Further ramifications of this model deserve study.

\end{document}